\documentclass[aps,prd,twocolumn,amsmath,amsfonts,amssymb,nofootinbib,floatfix]{revtex4}
\usepackage{graphicx,bm}
\usepackage{setspace,pstricks,color}
\definecolor{rossoCP3}{cmyk}{0,.88,.77,.40}

\begin{document}
\title{\Large {\color{rossoCP3}Mixed dark matter from technicolor}  }
\author{Alexander {\sc Belyaev}} \email{a.belyaev@phys.soton.ac.uk}
\affiliation{NExT Institute:
  School of Physics \& Astronomy, Univ. of Southampton, UK \\
  Particle Physics Department, Rutherford Appleton Laboratory, UK}
\author{Mads T. {\sc Frandsen}} \author{Subir {\sc Sarkar}}
\email{m.frandsen1@physics.ox.ac.uk} \affiliation{Rudolf Peierls
  Centre for Theoretical Physics, University of Oxford, UK}
\author{Francesco {\sc Sannino}} \email{sannino@cp3.sdu.dk}
\affiliation{CP$^\mathbf 3$-Origins, University of Southern Denmark,
  Odense M, DK}

\begin{abstract}
  We study natural composite cold dark matter candidates which are
  pseudo Nambu-Goldstone bosons (pNGB) in models of dynamical
  electroweak symmetry breaking. Some of these can have a significant
  thermal relic abundance, while others must be mainly asymmetric dark
  matter. By considering the thermal abundance alone we find a lower
  bound of $m_W$ on the pNGB mass when the (composite) Higgs is
  heavier than 115 GeV. Being pNGBs, the dark matter candidates are in
  general light enough to be produced at the LHC.
\end{abstract}
{\it CP$^3$- Origins: 2010-20}
\maketitle


\section{Introduction}

A second strongly coupled sector in Nature akin to QCD is a likely
possibility. The new strong interaction may naturally break the
electroweak (EW) symmetry through the formation of a chiral
condensate, thus making the Standard Model (SM) Higgs a composite
particle. Models of this type are called `technicolor' (TC)
\cite{Weinberg:1979bn,Susskind:1978ms} and several new variants have
been proposed recently \cite{Sannino:2004qp,
  Hong:2004td,Dietrich:2005wk,Dietrich:2005jn,Gudnason:2006mk,Ryttov:2008xe,Frandsen:2009fs,Frandsen:2009mi,Antipin:2009ks}
with interesting dynamics relevant for collider phenomenology
\cite{Foadi:2007ue,Belyaev:2008yj,Antola:2009wq,Antipin:2010it} as
well as cosmology
\cite{Nussinov:1985xr,Barr:1990ca,Bagnasco:1993st,Gudnason:2006ug,Gudnason:2006yj,Kainulainen:2006wq,Kouvaris:2007iq,Kouvaris:2007ay,Khlopov:2007ic,Khlopov:2008ty,Kouvaris:2008hc,Belotsky:2008vh,Cline:2008hr,Nardi:2008ix,Foadi:2008qv,Jarvinen:2009wr,Frandsen:2009mi,Jarvinen:2009mh,Kainulainen:2009rb,Kainulainen:2010pk,Frandsen:2010yj,Kouvaris:2010vv}.
A review of these models and the phase diagram of strongly coupled
theories can be found in Ref.~\cite{Sannino:2009za}. A relevant point
is that the technicolor dynamics is strongly modified by the new
interactions necessary to give masses to SM fermions
\cite{Fukano:2010yv} and the interplay between these two sectors leads
to an entirely new class of models, constraints on which were
discussed in Ref.~\cite{Chivukula:2010tn}.

We discuss different possibilities for dark matter (DM) candidates
within this rich framework and show that some of these composite
states can be thermal relics while being sufficiently light to be
produced at the LHC.

We call dark matter candidates composed of technicolor fields
`technicolor interacting massive particles' (TIMPs) and focus on those
which are pseudo Nambu-Goldstone bosons (pNGB). Our analysis is
general since we use a low energy effective description for the TIMPS
which can easily be adapted for specific models. We will discuss some
of these models
\cite{Gudnason:2006yj,Ryttov:2008xe,Foadi:2008qv,Frandsen:2009mi}
which provide particularly interesting candidates for dark matter.

\section{The simplest TIMPs from Partially Gauged Technicolor}

An interesting class of TIMPs arise from {\it partially gauged
  technicolor} models \cite{Dietrich:2005jn,Dietrich:2006cm} in which
only part of the TC group is gauged under the EW interactions. The EW
gauged technifermions are organized in doublets in the usual way while
the other technifermions are collectively denoted $\lambda^f$, with
$f$ counting these flavors only:
\begin{align}
\label{PGTC}
 Q_L = \begin{pmatrix}U_L \\ D_L \end{pmatrix} , U_R \
, \ D_R \ ; \quad \lambda^f 
\end{align}
These models were introduced originally in order to yield the smallest
na\"ive EW $S$ parameter, while still being able to achieve walking
dynamics.~\footnote{The na\"ive $S$-parameter from a loop of
  technifermions counts the number of fermion doublets transforming
  under weak $SU(2)_{\rm L}$, while walking dynamics is required to
  reduce non-perturbative contributions to the full $S$-parameter. The
  na\"ive $S$-parameter has recently been conjectured to be the
  absolute lower bound of the full $S$-parameter
  \cite{Sannino:2010ca,Sannino:2010fh}, making the TC models presented
  here optimal with respect to satisfying the LEP precision data.} The
non-minimal flavor symmetry of the resulting model allows for a number
of light states accessible at colliders. A similar scenario is
envisaged in so-called {\it conformal technicolor}
\cite{Luty:2004ye,Galloway:2010bp}. The technifermions not gauged
under the EW interactions essentially constitute a strongly
interacting {\it hidden sector}.

To be specific we consider a scalar TIMP, $\phi \sim \lambda \lambda$,
made of the SM gauge singlet technifermions $\lambda^f$ and possessing
a global $U(1)$ symmetry protecting the lightest state against
decay. Moreover we take $\phi $ to be a pNGB from the breaking of
chiral symmetry in the hidden sector, which leaves this $U(1)$
unbroken. This constitutes the simplest type of TIMP from the point of
view of its interactions so we study this first and consider later
TIMPs with constituents charged under the SM. An explicit model of
partially gauged technicolor featuring this kind of TIMP is `ultra
minimal technicolor' (UMT) \cite{Ryttov:2008xe}.  We therefore
identify our DM candidate with a complex scalar $\phi$, singlet under
SM interactions and charged under a new $U(1)$ symmetry ({\it not} the
usual technibaryon symmetry), which makes it stable.

In addition to the TIMP we consider a light (composite) Higgs boson. A
general effective Lagrangian to describe this situation is presented
below and can be derived, for any specific model, from the UMT
Lagrangian \cite{Ryttov:2008xe}. At low energies we can describe the
interactions of $\phi$ through a chiral Lagrangian. The (composite)
Higgs $H$ couples to the TIMP as:
\begin{eqnarray}
\label{Basic Lagrangian}
\mathcal{L} &=& \partial_\mu\phi^\ast\partial_\mu\phi - m_\phi^2 \phi^\ast \phi 
+ \frac{d_1}{\Lambda}\ H\partial_\mu\phi^\ast\partial_\mu\phi \\ \nonumber
&+& \frac{d_2}{\Lambda} m_\phi^2 H \phi^\ast \phi 
+  \frac{d_3}{2\Lambda^2} H^2 \partial_\mu\phi^\ast\partial_\mu\phi
+ \frac{d_4}{2\Lambda^2} m_\phi^2 H^2 \phi^\ast \phi .
\end{eqnarray}
The interactions between technihadrons such as $\phi$ made of EW
singlet constituents and states with EW charged constituents
(e.g. $H$) are due mainly to TC dynamics and, as such, the couplings
between these two sectors are {\em not} suppressed
\cite{Ryttov:2008xe}. However, since $\phi$ is a pNGB it must have
either derivative couplings or the non-derivative couplings must
vanish in the limit $m_\phi \to 0$. The mass $m_\phi$ is assumed to
come from interactions beyond the TC sector, e.g. `extended
technicolor' (ETC) \cite{Eichten:1979ah,Appelquist:2002me} which can
provide masses for the TC Nambu-Goldstone bosons, as well as for SM
fermions.  The couplings $d_1, ... , d_4$ are dimensionless and
expected to be of $O(1)$, while $\Lambda$ is the scale $\Lambda \sim
4\pi F_{\pi}$ below which the derivative expansion is sensible.

%
%

We emphasize the differences between a composite scalar TIMP and {\em
  fundamental} scalar dark matter considered earlier
\cite{McDonald:1993ex,Cirelli:2005uq,Andreas:2008xy}: i) The $U(1)$ is
natural, i.e. it is identified with a global symmetry (not necessarily
the technibaryon one), ii) Its pNGB nature makes the DM candidate
naturally light with respect to the EW scale and influences the
structure of its couplings, iii) Compositeness requires the presence
around the EW scale of spin-1 resonances in addition to the TIMP and
the (composite) Higgs; their interplay can lead to striking collider
signatures \cite{Foadi:2008qv,Frandsen:2009mi}.

\section{Thermal versus asymmetric dark matter}

When the TIMP is a composite state made of particles charged under the
EW interactions it becomes a good candidate for {\em asymmetric} dark
matter, i.e. its present abundance is due to a relic asymmetry between
the particle and its antiparticle, just as for baryons. This has been
the case usually considered when discussing TC DM candidates
\cite{Nussinov:1985xr,Chivukula:1989qb} since the technibaryon
self-annihilation cross-section, obtained by scaling the
proton-antiproton annihilation cross-section up to the EW scale, is
high enough to essentially erase any symmetric thermal relic
abundance. Hence an asymmetry between technibaryons and
anti-technibaryons is invoked, especially as this can be generated
quite naturally in the same manner as for baryons. However, scaling up
the proton-antiproton annihilation cross-section is not applicable to
generic TIMPs, in particular not to pNGBs, hence they may have an
interesting {\em symmetric} (thermal) relic abundance.

Let us solve for the thermal relic abundance of TIMPs $\phi$ with
singlet constituents using the Boltzman continuity equation
\cite{Lee:1977ua,Vysotsky:1977pe}:
\begin{equation}
\label{continuity}
 \frac{{\rm d}}{{\rm d} t} (n_{\phi} R^3) = 
-{\langle\sigma_{\rm ann}\,v\rangle}\left[n_\phi^2 - (n_\phi^{\rm eq})^2\right] R^3\ ,
\end{equation}
where $R$ is the cosmological scale-factor, $n_\phi$ the TIMP number
density and $\sigma_{\rm ann}$ the TIMP-antiTIMP annihilation
cross-section (given in Appendix \ref{Acs} along with the relevant
interaction vertices). From the Lagrangian (\ref{Basic Lagrangian}) we
see that annihilations proceed via the (composite) Higgs into SM
fermions and gauge bosons pairs, as well as into a pair of (composite)
Higgs particles. As discussed in
Refs.\cite{Griest:1986yu,Gondolo:1990dk,Abel:1992ts}, we can rewrite
the continuity equation (\ref{continuity}) in terms of the
dimensionless quantities $Y \equiv n_\phi/s$, $Y^{\rm eq} \equiv
n^{\rm eq}/s$, and $x \equiv m_\phi/T$, where $s \equiv g_s T^3$ is
the specific entropy determining the value of the adiabat $RT$:
\begin{eqnarray}
\label{Yseqn}
\frac{{\rm d}Y}{{\rm d} x} & = & \lambda x^{-2} \left[
  (Y^{\rm eq})^{2}
  - Y^{2} \right] , \\ \nonumber
\mbox{where,} \quad
\lambda &\equiv& \left(\frac{g_s^4}{180\pi}\right)^{1/6}\, 
m_\phi\,m_{\rm P}\,{\langle\sigma_{\rm ann}\,v \rangle}\,g_\rho^{1/2}\ , 
\end{eqnarray}
and $g_\rho \equiv \rho/T^4$ counts the number of relativistic
degrees of freedom contributing to the energy density, which
determines the Hubble expansion rate $\dot{R}/R$. The values of
$g_\rho(T)$ and $g_s(T)$ have been computed in the SM
\cite{Srednicki:1988ce} and are modified to account for the additional
particle content of TC models.

In the hot early universe, the particle abundance initially tracks its
equilibrium value but when the temperature falls below its mass and it
becomes non-relativistic, its equilibrium abundance falls
exponentially due to the Boltzmann factor. Hence so does the
annihilation rate, eventually becoming sufficiently small that the
(comoving) particle abundance becomes constant.  Defining the
parameter $\Delta \equiv (Y-Y^{\rm eq})/Y^{\rm eq}$, the freeze-out
temperature is given by \cite{Griest:1986yu}:
\begin{eqnarray}
\label{freeze-out temp}
&x_{\rm fr} = \frac{1}{b_s}\ln\left[\Delta_{\rm fr}(2 
+ \Delta_{\rm fr})\delta_s\right]
- \frac{1}{2b_s}\ln\left[\frac{1}{b_s}\ln(\Delta_{\rm fr}(2 
  + \Delta_{\rm fr})\delta_s)\right] \nonumber \\
 &\mbox{where,}  \quad
b_s \equiv \left(\frac{2\pi^2 g_s}{45}\right)^{1/3}, \quad
\delta_s = \frac{g}{(2\pi)^{3/2}}b_s^{-5/2} \lambda,
\end{eqnarray}
and $g$ counts the internal degrees of freedom, e.g. $g=2$ for the
TIMP. This gives a good match to the exact numerical solution of
Eq.(\ref{Yseqn}) for the choice $\Delta_{\rm fr} = 1.5$ which
corresponds to the epoch when the annihilation rate, $n_{\phi}^{\rm
  eq}{\langle\sigma_{\rm ann}\,v\rangle}$, equals the logarithmic rate
of change of the particle abundance itself: ${\rm d}\ln n^{\rm
  eq}/{\rm d}t = x_{\rm fr}\ \dot{R}/R$ \cite{Lee:1977ua}. Note that
the usual criterion of equating the annihilation rate to the Hubble
expansion rate $\dot{R}/R$ would give an {\em erroneous} answer when
there is an asymmetry \cite{Griest:1986yu}.

We calculate the TIMP freeze out parameter $x_{\rm fr}$ as a function
of the TIMP mass $m_{\phi}$ taking $\Lambda=1$ TeV, for three values
of the (composite) Higgs mass $m_H=250, 500, 1000$~GeV. We also take
the dimensionless effective couplings to the (composite) Higgs to be
of ${\cal O}(1)$ and define $d_{12}\equiv d_1+d_2$, $d_{34}\equiv
d_3+d_4$, since at low energies the $d_1$ and $d_2$ terms contribute
very nearly equally, as do the $d_3$ and $d_4$ terms. There are spikes
in $x_{\rm fr}$ at the (composite) Higgs resonance when $2 m_\phi =
m_H$, however the simple approximation above is not reliable near such
a resonance \cite{Griest:1990kh} and we must then solve the full
continuity equation including the (composite) Higgs width. We do this
using the programme {\tt MicrOMEGAs}
\cite{Belanger:2001fz,Belanger:2008sj,Belanger:2010gh} which computes
the full annihilation cross-section of the model using {\tt CalcHEP}
\cite{Pukhov:2004ca}. We also use {\tt LanHEP} \cite{Semenov:2008jy}
for the model implementation.

\begin{figure}[htp!]
\includegraphics[width=\columnwidth]{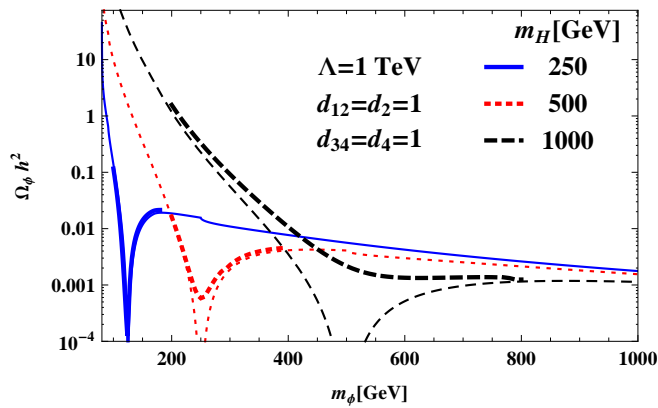}
\caption{The relic TIMP ($\phi$) abundance vs. its mass. The thick
  lines show the {\tt Micromegas} computation taking into account the
  (composite) Higgs decay width.}
\label{fig:omegah21}
\end{figure}

After freeze-out, only annihilations are important since the
temperature is now too low for the inverse creations to proceed; the
asymptotic abundance is then:
\begin{eqnarray}
\label{asymptotic density}
Y_{\infty} \equiv Y (t \rightarrow \infty)
 \sim \frac{ x_{\rm fr}}{ \lambda_{\rm fr}}  \ .
\end{eqnarray}
The resulting cosmological energy density of relic TIMPs is shown in
Fig.~\ref{fig:omegah21}. We have checked explicitly with the numerical
code that the contributions from $d_1$ and $d_2$ terms are (very
nearly) identical, as are the contributions from
$d_3$ and $d_4$ terms.

\begin{figure}[htp!]
  \includegraphics[width=\columnwidth]{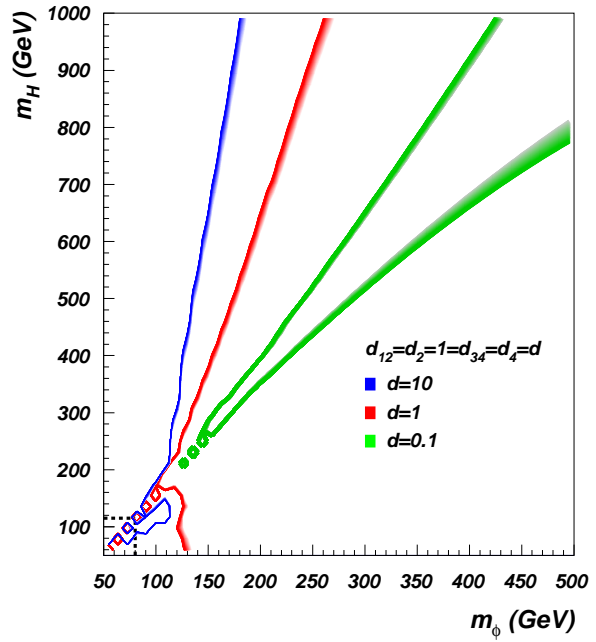}
  \caption{Regions corresponding to $\Omega h^2=0.11\pm 0.01$ for the
    relic TIMP ($\phi$) abundance in the (composite) Higgs vs. TIMP
    mass plane. The dashed box shows that given $m_H > 115$ GeV, we
    require $m_\phi > m_W$ for TIMPs to be dark
    matter.}\label{fig:omegah2025}
\end{figure}

Fig.~\ref{fig:omegah2025} shows the region in the (composite) Higgs
versus TIMP mass plane where the TIMP relic abundance matches the DM
abundance $\Omega h^2=0.11\pm 0.01$ ($2\sigma$) inferred from WMAP-7
\cite{Larson:2010gs}.

From Figs.~\ref{fig:omegah21} and \ref{fig:omegah2025} we observe
first that the relic energy density drops significantly for $m_\phi \sim 2
m_H$ as expected (because $\phi$ can then decay resonantly through the
Higgs) and, second, that if $m_H$ is greater than about 115 GeV then
for TIMPs to be dark matter requires $m_\phi > m_W$.

As discussed below, in the presence of an asymmetry the total relic
abundance always increases relative to the same model with no
asymmetry, so $m_W$ provides a general {\em lower} bound for the mass
of the pNGB TIMPs we consider. It follows that in the interesting
region $m_W \lesssim m_\phi \lesssim 1$ TeV, {\em symmetric} relic
TIMPs with singlet constituents could make up a significant fraction
of the dark matter in the universe. However when the TIMP is heavier
than about a TeV, the strength of the interaction is similar to that
of an ordinary (scalar) technibaryon, and a relic abundance large
enough to account for dark matter now does require an initial {\em
  asymmetry} similar to that of baryons as discussed earlier
\cite{Nussinov:1985xr,Barr:1990ca}.  We note that recently a different
type of QCD-like pions (which do not carry a $U(1)$ quantum number)
were also considered as dark matter candidates
\cite{Hur:2007uz,Bai:2010qg}.

\subsection{Adding an asymmetry}

\begin{figure}[htp!]
  {\includegraphics[width=\columnwidth]{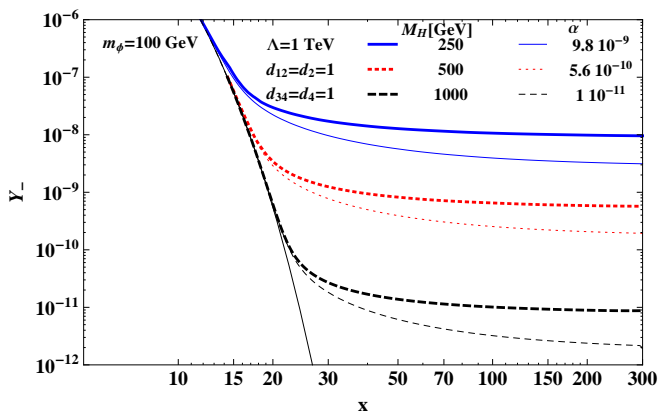}}
  \caption{TIMP ($\phi$) abundance when there is no asymmetry (thick
    lines), compared with the abundance of the {\em minority} species
    $Y_-$ (thin lines) when an asymmetry $\alpha$ is present.}
\label{fig:asymmetry}
\end{figure}

To study the relic abundance in the presence of both a thermal
component and an initial asymmetry we follow Ref.\cite{Griest:1986yu}
and define the asymmetry as $\alpha=(Y_+ - Y_-)/2$ where $Y_\pm$ are
the abundances of the majority and minority species (TIMP and
anti-TIMP) respectively. The abundance in thermal and chemical
equilibrium is:
\begin{eqnarray}
Y_-^{\rm eq} = 
{\rm e}^{-\mu/T} Y^{\rm eq} \sim e^{-\mu/T} g 
\left(\frac{x}{2 \pi b_s} \right)^{3/2} {\rm e}^{-b_s x}\ ,
\end{eqnarray}
where $\mu$ is the chemical potential. 
The continuity equation in the presence of an asymmetry is
\cite{Griest:1986yu}:
\begin{eqnarray}
\label{Yseqnwasym}
\frac{{\rm d}Y_-}{{\rm d}x} & = & \lambda x^{-2} \left[
  Y_{-}^{\rm eq}(Y_{-}^{\rm eq}+2 \alpha)
  - Y_{-}(Y_{-} + 2\alpha) \right] ,
\end{eqnarray}
and the total asymptotic abundance of TIMPs and anti-TIMPs is:
\begin{eqnarray}
\label{asymptotic density w asymmetry}
\Omega_{\phi} h^2 = 5.5 \times 10^8(Y_{\rm -\infty} 
 + \alpha) \frac{m_\phi}{\rm GeV}. 
\end{eqnarray}
In Fig.~\ref{fig:asymmetry} we show the minority species abundance
$Y_-$ as a function of $x \equiv m_\phi/T$ for $m_\phi=$ 100 GeV and
$\alpha=9.8 \times 10^{-9}, 5.6 \times 10^{-10}, 10^{-11}$, taking the
Higgs mass to be $m_H = 250, 500, 1000$~GeV.

We see from the figure that the {\em symmetric} component of the relic
abundance is $\sim 10\%$ of the asymmetric component, when the
asymmetry is comparable to the would-be symmetric abundance (in the
absence of an asymmetry). In the limit where $\alpha \ll Y_{\infty}$,
the symmetric component is unchanged and the asymmetry provides a
small addition to the total relic abundance.  When $\alpha\gtrsim
Y_\infty$, the abundance of the minority species is exponentially
suppressed in $\alpha$ and provides a negligible addition to the
asymmetric component. Adding an asymmetry will always {\em increase}
the relic abundance relative to its value in the same model in the
absence of an asymmetry. This implies a non-trivial constraint on the
mass of TIMPs with uncharged constituents discussed above, such as
appear in e.g. the UMT model. The constraint is non-trivial since such
states would evade direct detection at colliders unless the
(composite) Higgs is very light \cite{Foadi:2008qv}, as well as the
direct detection experiments discussed below.

\begin{figure}[htp!]
  \includegraphics[width=\columnwidth]{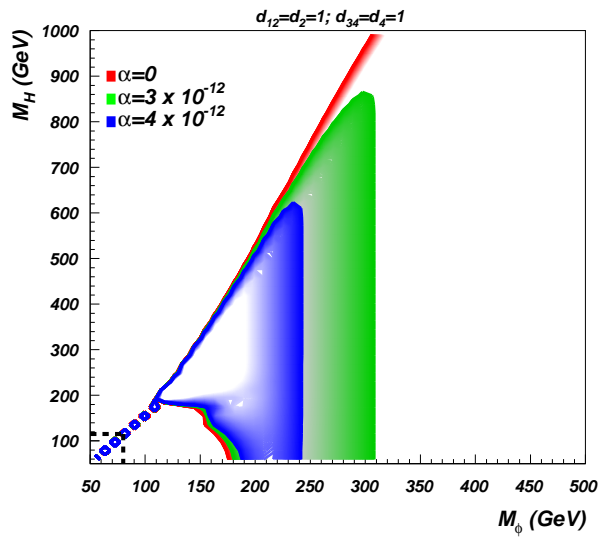}
  \caption{Regions in the (composite) Higgs versus TIMP mass plane
    corresponding to $\Omega h^2=0.11\pm 0.01$ for the TIMP ($\phi$)
    relic abundance, for different values of the relic asymmetry
    $\alpha$. The dashed box shows that given $m_H > 115$ GeV, we
    require $m_\phi > m_W$ for TIMPs to be dark matter.}
\label{fig:omegah2025_asym}
\end{figure}

Fig.~\ref{fig:omegah2025_asym} shows the contour in the $m_H-m_\phi$
plane where the relic abundance of $\phi$ agrees with the DM abundance
inferred from WMAP-7 \cite{Larson:2010gs} for a fixed value of the $d$
coefficients. Just as in Figs.~\ref{fig:omegah21} and
\ref{fig:omegah2025} we observe that if $m_H$ is greater than $\sim
115$ GeV we require $m_\phi > m_W$ to avoid an excessive TIMP relic
abundance.  However, due to the asymmetry, for TIMP masses above $m_W$
there is now a much broader range of $m_H$ and $m_\phi$ where the
relic abundance of $\phi$ matches the observed DM abundance.

\subsection{TIMPs with charged constituents}

We consider now pNGB TIMPs with charged constituents of the form
$T\sim U D$ arising from the TC sector carrying EW interactions (see
Eq.~\ref{PGTC}). These states carry an $U(1)$ quantum number which
makes them stable and it is natural to identify this global symmetry
with the technibaryon number. Such particles arise generally in TC
models with the technifermions transforming in either real or
pseudo-real representations of the gauge group. Explicit examples are
furnished again in the UMT scheme in which the composite $T$ is a SM
singlet and the `orthogonal minimal technicolor' (OMT) model in which
the $T$ state is the isospin-0 component of a complex triplet
\cite{Frandsen:2009mi}.

We demonstrate that similarly to the case of TIMPs with neutral
constituents, TIMPs with charged constituents have a significant
symmetric component in only a small region of parameter
space. However, as opposed to the TIMPs with SM neutral constituents
this region is essentially independent of the parameters of the
(composite) Higgs interactions.  In addition to the interactions
investigated above, the scalar TIMPs containing charged constituents
will also have an effective interaction with the photon, due to a
non-zero electromagnetic charge radius of $T$
\cite{Bagnasco:1993st,Foadi:2008qv}:
\begin{eqnarray} 
  \mathcal{L}_B = i e \frac{d_B}{\Lambda^2} T^* 
  \overleftrightarrow{\partial_\mu} T \, \partial_{\nu}F^{\mu\nu} \ .
\label{eq:db}
\end{eqnarray}
The corresponding charge radius of the TIMP is $r_T \sim
\sqrt{d_B}/\Lambda$.  For our choice $\Lambda =1$ TeV we consider the
range $|d_B|=0$ to $0.3$ while for a higher cut-off $\Lambda$, a
larger $d_B\sim O(1)$ would be expected.

In case of the TIMP $T$ there are also contact interactions with two
SM vector bosons $V$, arising from the kinetic term of the chiral
Lagrangian, which can significantly affect the symmetric relic
density. In general these can be written as
\begin{eqnarray}
\label{contactint}
L_{VV} = \frac{1}{2} T^{\ast} T V_{\mu} V^{\mu}\ {\rm Tr} &[&[\Lambda_S, 
[\Lambda_S, X_T]] X_{T^*} \\ \nonumber
&-& [\Lambda_S, [\Lambda_S, X_T]] X_{T^*}] ,
\end{eqnarray}
where $X_{T}$ is the generator of the broken (techni) flavor direction
corresponding to the TIMP and $\Lambda_S$ are the, appropriately
normalized, EW generators imbedded in the TC chiral group
\cite{Preskill:1980mz,Peskin:1980gc,Chadha:1981rw}.
The resulting $TTWW$ and $TTZZ$ contact interactions (assuming that
the EW symmetry is broken already) are:
\begin{equation}
  L_{WW,ZZ} = - \frac{T^{\ast}T}{2} {\rm Tr} 
  \left[ d_W \, W_{\mu}W^{\mu} + d_Z  \, Z_{\mu}Z^{\mu} \right] \ ,
\label{eq:dw}
\end{equation}
with $d_W = g^2$ and $d_Z = (g^2 +{g^{\prime}}^2)/2$ for the TIMPs $T$
of both the UMT and the OMT models. 

\begin{figure}[htp!]
\includegraphics[width=\columnwidth]{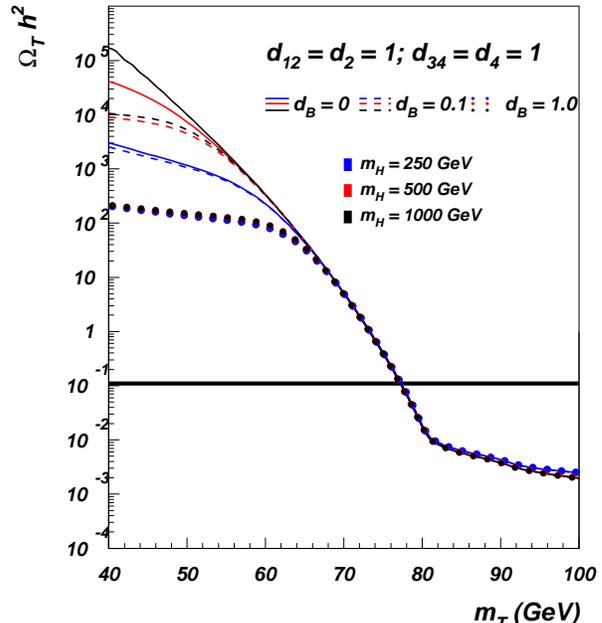}\\ 
\vspace*{-0.8cm}
\caption{The TIMP ($T$) relic energy density as a function of its mass
  when $W$ and $Z$ contact interactions are present, calculated taking
  into account their off-shell decays. The horizontal band corresponds
  to $\Omega h^2 = 0.11 \pm 0.01$.}
\label{fig:xfr_WZ}
\end{figure}

In Fig.~\ref{fig:xfr_WZ} we display the effect of the $W$ and $Z$
contact interactions, as well as of the charge radius interaction, on
the thermal relic abundance. It is seen that for $m_T$
significantly below $m_W$, the interactions in Eq.(\ref{eq:db})
do affect TIMP annihilations significantly, even so the TIMP relic
abundance is too large. As $m_T$ increases towards $m_W$,
annihilations due to the $TTVV$ interactions in Eq.(\ref{eq:dw}) begin
to dominate and reduce the TIMP relic abundance to the observationally
acceptable level only when the TIMP is a few GeV lighter than the $W$.


\begin{figure}[htp!]
{\includegraphics[width=\columnwidth]{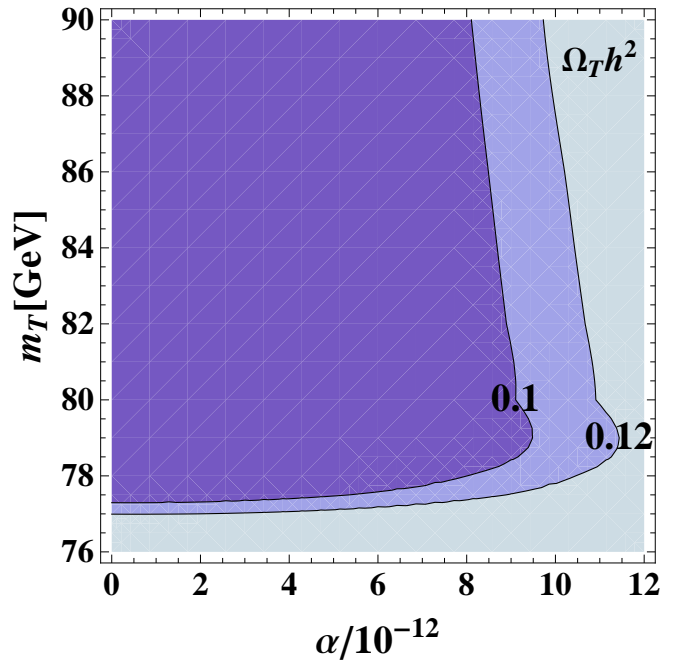}}
\vspace*{-0.3cm}
\caption{The region in the asymmetry ($\alpha$) vs. TIMP ($T$) mass
  plane which yields the marked relic energy density (consistent with
  WMAP) when $W$ and $Z$ contact interactions are present.}
 \label{fig:omegah2_WZ_asym}
\vspace*{-0.5cm}
\end{figure}

Again, once we include an asymmetry, the range of TIMP masses where a
cosmologically acceptable {\em symmetric} relic abundance is achieved,
can be much broader. This is demonstrated in
Fig.~\ref{fig:omegah2_WZ_asym} which shows the contours of $\Omega
h^2$ for different values of the asymmetry and $m_T$.  There is a
cross-over in the small TIMP mass region mentioned above, from just
below $m_W$ where the relic abundance is dominated by the asymmetry,
to just above $m_W$ where it is dominated by the symmetric component.


\section{Direct Detection}
\label{Direct Detection}

Direct detection of the TIMPs $\phi$ with SM singlet constituents will
be challenging. The exchange of the (composite) Higgs leading to a
scattering cross-section on nuclei is the most relevant interaction
here, and we will assume that the (composite) Higgs couples to SM
fermions with ordinary Yukawa couplings. (In fact this represents an
upper bound and so the actual cross-section could be lower). Again,
since the TIMP is a pNGB the couplings to the Higgs are suppressed at
low masses. The TIMP nucleon scattering cross-section from the
(composite) Higgs exchange is given by:
\begin{equation}
\sigma_{\textrm{nucleon}}^H = \frac{\mu^2}{2\pi} 
\left[\frac{d_H  f m_N}{m^2_H m_\phi v}\right]^2,  \quad
d_H = (d_1 + d_2)\frac{m_\phi^2}{\Lambda},
\end{equation}
where $\mu$ is the nucleon-TIMP reduced mass, $v$ the electroweak vev
and $f$ parameterizes the (composite) Higgs-nucleon coupling. We refer
to Refs.\cite{Ohki:2008ff,Giedt:2009mr} for recent discussions on the
strange quark contribution to $f$ which we take to be $f=0.3$.

\begin{figure}[htp!]
\vspace*{-0.8cm}
\includegraphics[width=\columnwidth]{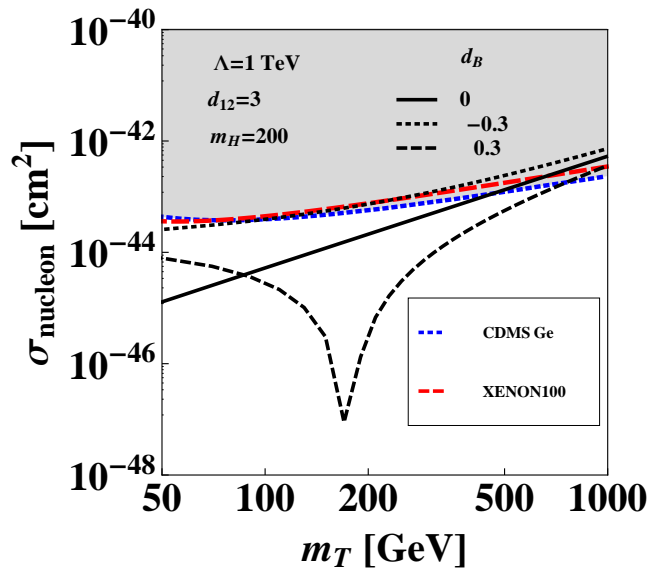}
\caption{ Direct detection cross-section (per nucleon) for TIMP
  scattering off nuclei.  The full line is for Higgs exchange only
  (with $m_H = 200$ GeV) while the short- and long-dashed lines show
  the additional effect of the charge radius operator (with $d_B=-0.3$
  and $d_B=+0.3$ respectively).  The shaded region is experimentally
  excluded.}
\label{fig:cdmsdm2}
\end{figure}

For TIMPs $T$ with charged constituents there is an additional
contribution to the scattering on nuclei via the charge radius
operator \cite{Bagnasco:1993st,Foadi:2008qv}
\begin{eqnarray}
  \sigma_{\textrm{p}}^{\gamma} &=& \frac{\mu^2}{4\pi} \left[\frac{ 8\pi \, 
  \alpha \, d_B}{\Lambda^2}\right]^2 \ .
\end{eqnarray}
To take into account the possible interference between the composite
Higgs and photon exchange \cite{Foadi:2008qv} we write an averaged
scalar cross-section per nucleon as
\begin{eqnarray}
\sigma_{\textrm{nucleon}} &\equiv& \frac{\mu^2}{4\pi A^2}(f_p Z+ f_n (A-Z))^2 
\\ \nonumber
\mbox{where:} \quad 
f_n &=& -\frac{\sqrt{2} d_H  f m_N}{m^2_H m_\phi v}, \quad f_p=f_n+\frac{ 8\pi \, 
  \alpha \, d_B}{\Lambda^2}.
\end{eqnarray}
The direct detection cross-section per nucleon as a function of the
TIMP mass is shown in Fig.~\ref{fig:cdmsdm2}, where we also indicate,
(following Ref.\cite{McCabe:2010zh}) the limits from the CDMS II
\cite{Ahmed:2009zw} and XENON-100 \cite{Aprile:2010um} experiments.

For Higgs exchange only ($d_B=0$) the direct detection cross-section
{\em increases} with TIMP mass since the TIMPs are pNGBs ({\em
  c.f.} Ref.~\cite{Foadi:2008qv} where the TIMPs were not derivatively
coupled to the (composite) Higgs).  However, the charge radius
interaction can significantly alter the
cross-section, by up to 2 orders of magnitude for $|d_B| \simeq 0.3$,
thus greatly affecting the discovery potential of direct detection
experiments.  For example, for $d_B=-0.3$ and $d_{12}=3$ the signals
from the TIMP $T$ with SM charged constituents would have been
observed already by CDMS II and XENON-100.  On the other hand, for {\it
  positive} values of $d_B$, there is a destructive interference
between (composite) Higgs and photon exchange, lowering the cross
section to $10^{-47}$~cm$^2$ at $m_T \simeq 200$~GeV for a particular
choice of the parameters.

\begin{figure}[htp!]
\includegraphics[width=\columnwidth]{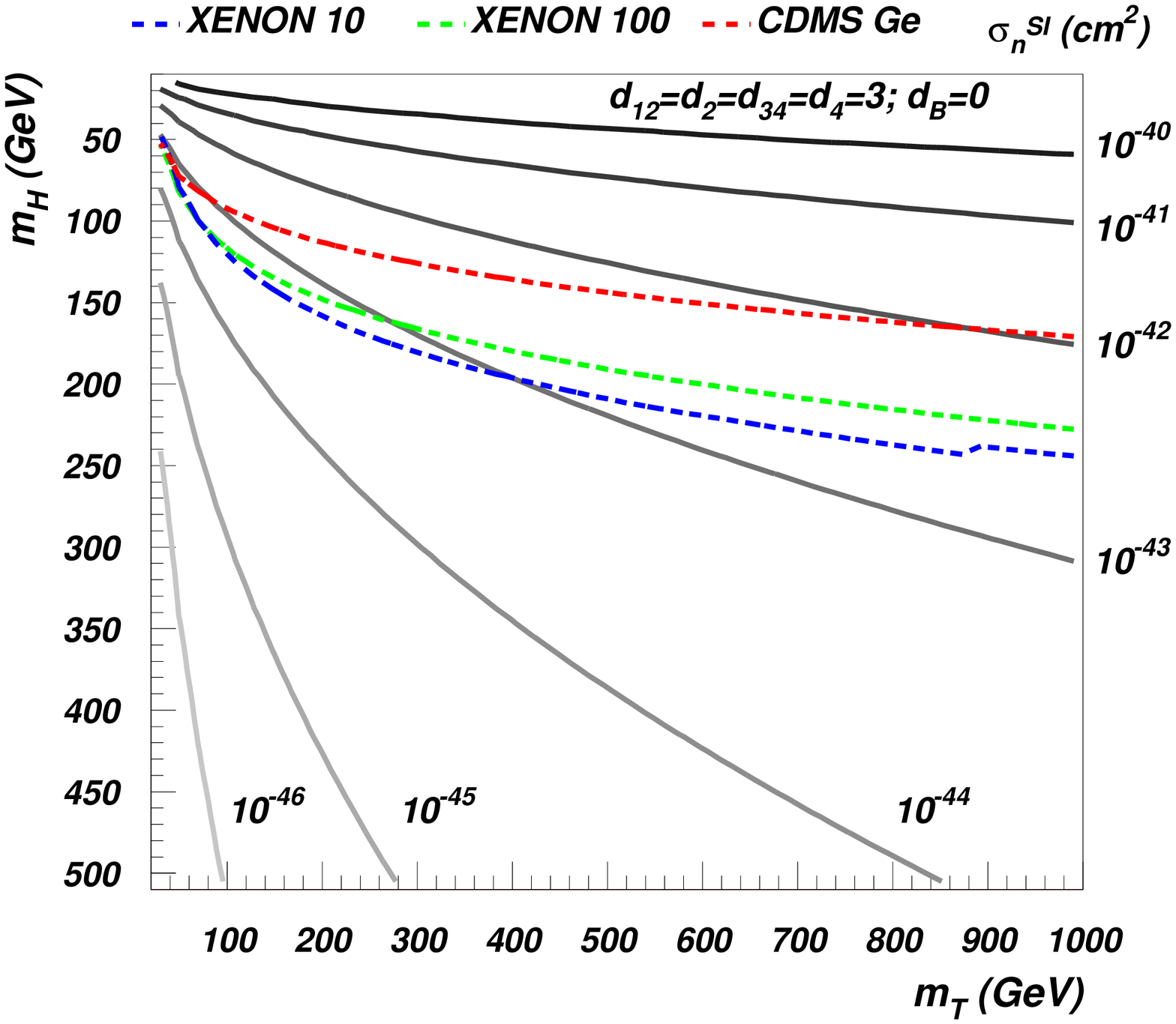}\\
\vskip -0.5cm
\includegraphics[width=\columnwidth]{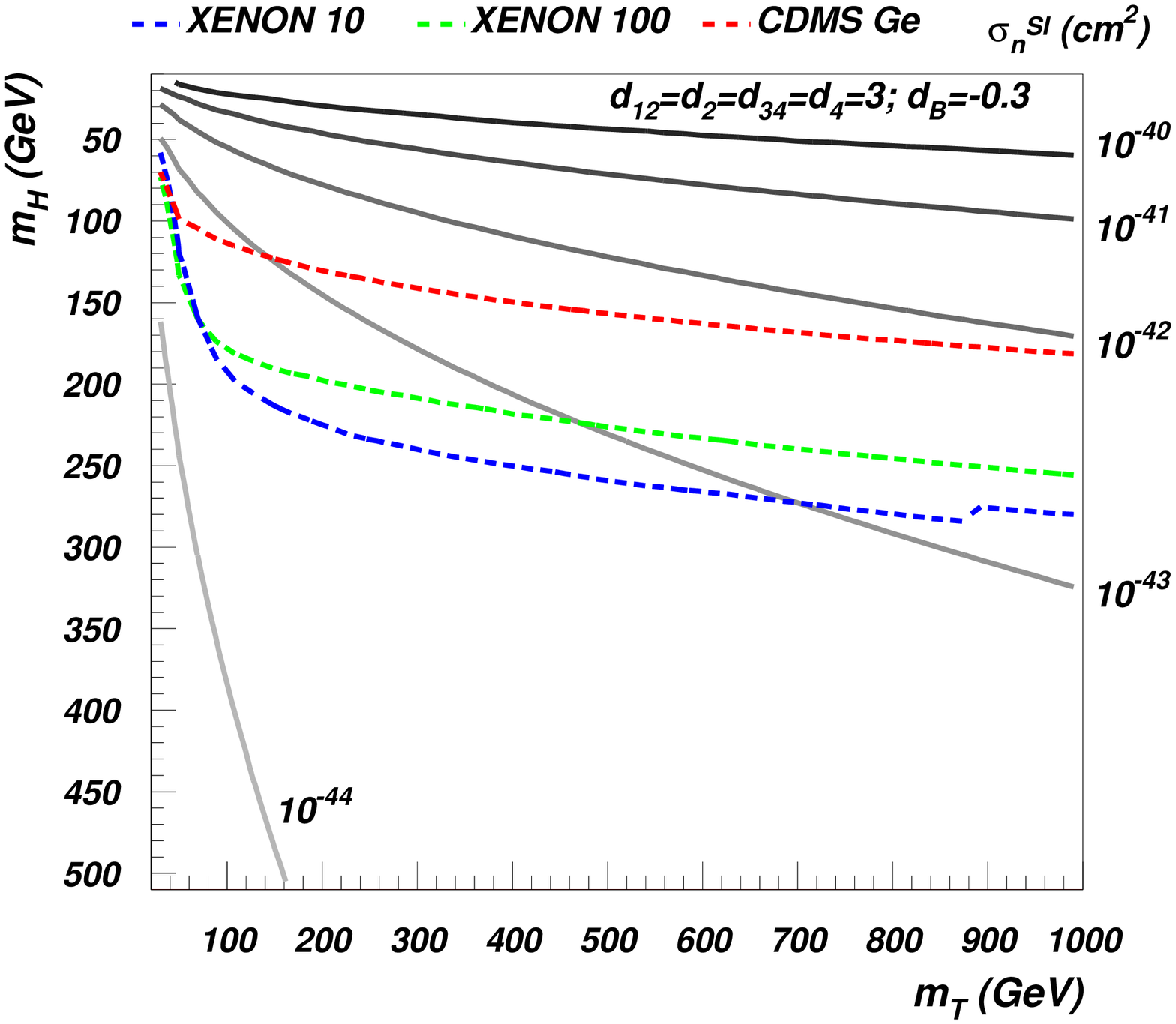}\\
\vskip -0.5cm
\includegraphics[width=\columnwidth]{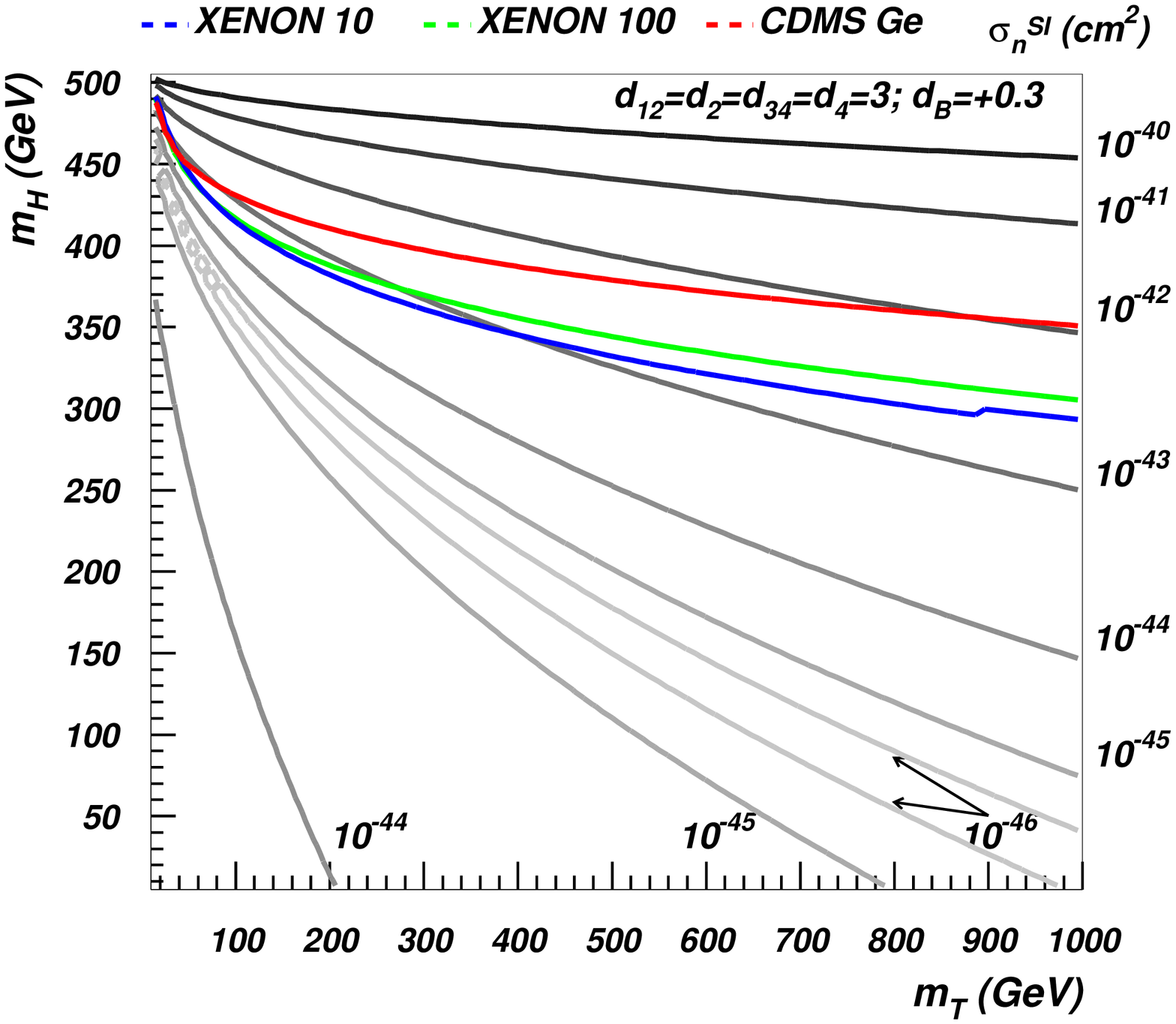}
\caption{Contours of the direct detection cross-section (per nucleon)
  for TIMP scattering off nuclei in the TIMP mass vs. (composite)
  Higgs mass plane. The top, middle and bottom frames are for
  $d_B=0,-0.3,+0.3$ respectively. The dashed curves show the upper
  limits from CDMS II (red), XENON-10 (blue) and XENON-100 (green).}
\label{fig:dd2d}
\end{figure}

Since the direct detection rate depends strongly on the (composite)
Higgs mass, we present in Fig.~\ref{fig:dd2d} results in the
$(M_H-m_T)$ plane for different values of the $d_B$ parameter.  The
figure also shows the exclusion limits from the XENON-10
\cite{Angle:2007uj}, CDMS II \cite{Ahmed:2009zw} and XENON-100
\cite{Aprile:2010um} experiments.  One can see that for $d_B=0$ and
$d_{12}=3$ (top frame), XENON-100 and CDMS II can cover essentially
the whole range of $m_T$ for $M_H$ below 150 GeV where the
cross-section always exceeds $10^{-42}$~cm$^2$.  Our results trivially
scale as $d_{12}^2$ for $d_B=0$.  Negative $d_B=-0.3$ significantly
enhances the cross-section (middle frame) while positive $d_B=0.3$
(bottom frame) brings the negative interference effect into play,
resulting in a deep valley.

\section{Conclusions}

We have shown that models of dynamical EW symmetry breaking can
provide symmetric (thermal) DM, as well as asymmetric (non-thermal)
DM. This is true in particular for {\it partially gauged technicolor}
\cite{Dietrich:2005jn,Dietrich:2006cm} which can satisfy constraints
from EW precision measurements.

From our analysis we conclude that for pNGB TIMPs: 

1) The TIMP cannot be significantly lighter than the $W$ if its relic
abundance is to be acceptable, unless the (composite) Higgs mass is
below 115 GeV --- this holds whether it is symmetric, asymmetric or a
combination.
    
2) If the TIMP is made of constituents charged under EW interactions,
it can be symmetric DM only in a narrow mass range close to $m_W$.
Above this mass an initial asymmetry is required for TIMPs to be dark
matter.

3) If the constituents of the TIMP are neutral with respect to the EW
interactions then it can be symmetric dark matter for a range of
masses tied to the (composite) Higgs mass. This does not exclude the
possibility that it has an asymmetry as well.

4) Direct detection of light pNGB TIMPs with neutral constituents is
challenging due to its mass suppressed couplings to the (composite)
Higgs.  However, for TIMPs with charged constituents there is an
additional charge radius interaction which, if sizeable, can bring
such TIMPs within the reach of current nuclear recoil detection
experiments.

Both types of TIMPs considered here -- with charged or neutral
constituents -- can co-exist and contribute to the dark matter (as in
e.g. the UMT model \cite{Ryttov:2008xe}). These models provide
interesting signals for direct dark matter detection experiments which
are already sensitive enough to exclude TIMPs in certain regions of
parameter space or even discover them in the near future.

\section*{Acknowledgements}

We wish to thank A. Pukhov for providing a modified version of {\tt
  MicrOMEGAs} (including an asymmetry in the solution of the
continuity equation), and C. McCabe for providing fits to direct
detection experimental data. MTF acknowledges a VKR Foundation
Fellowship. SS acknowledges support by the EU Marie Curie Network
``UniverseNet'' (HPRN-CT-2006-035863).

\appendix
\section{Annihilation cross-section}
\label{Acs}
The relevant vertex factors at low energies (if we keep only the
light (composite) Higgs in the spectrum) are:
{\small 
\begin{eqnarray}
  \phi^\ast \phi H &:& i \left(\frac{p_{\phi^\ast} p_\phi}{\Lambda} d_1 
  + d_2 \frac{m_\phi^2}{\Lambda}\right) \to i d_{12} \frac{m_\phi^2}{\Lambda}
  \nonumber \\
  \phi^\ast \phi H H &:& i \left(\frac{p_{\phi^\ast} p_\phi}{\Lambda^2} d_3 
  + d_4\frac{m_\phi^2}{\Lambda^2}\right) \to i d_{34} \frac{m_\phi^2}{\Lambda^2}
\end{eqnarray}
} where $d_{12}=d_1+d_2$ and $d_{34}=d_3+ d_4$ are the only
independent parameters at low energies. Hence the (composite) Higgs
mediated contributions to the cms annihilation cross-section
$\langle\sigma v_{\rm rel}\rangle$ in the limit $v_{\rm rel} \to 0$
are: \newline $\underline{\phi \phi^{\ast} \rightarrow HH}:$ {\small
\begin{eqnarray}
 \frac{1}{64 \pi m_{\phi}^{2}} \left[\frac{3 d_{12} m_H^2 m_\phi^2 }{v \Lambda 
    \left(4 m_\phi^2 - m_H^2\right)} - \frac{2 d_{12}^2  m_\phi^4}
    {\Lambda^2 \left(m_H^2 - 2m_\phi^2\right)} +
   \frac{d_{34} m_\phi^2 }{\Lambda^2}\right]^2 \nonumber \\
  \times \left(1 - \frac{m_H^2}{m_\phi^2}\right)^{1/2} ,
  \end{eqnarray}
}
\newline 
$\underline{\phi \phi^\ast \rightarrow W^+ W^-}:$
{\small 
\begin{eqnarray}
  2 \left[1 + \frac{1}{2}\left(1 - \frac{2 m_\phi^2}{m_W^2}\right)^2\right] 
  \frac{d_{12}^2 m_\phi^2 m_W^4}{8 \pi v^2 \Lambda^2 
    \left[\left(4m_\phi^2 - m_h^2\right)^2 + m_h^2\Gamma_h^2\right]} \nonumber \\
  \times \left(1-\frac{m_W^2}{m_\phi^2}\right)^{1/2} 
\end{eqnarray}
}
\newline 
$\underline{\phi \phi^\ast \rightarrow ZZ}:$
{\small 
\begin{eqnarray}
  2\left[1 + \frac{1}{2}\left(1 - \frac{2 m_\phi^2}{m_Z^2}\right)^2\right]
  \frac{d_{12}^2 m_\phi^2 m_Z^4}{16\pi v^2 \Lambda^2 \left[\left(4 m_\phi^2
        - m_h^2\right)^2 + m_h^2 \Gamma_h^2\right]} \nonumber \\
  \times \left(1 - \frac{m_Z^2}{m_\phi^2}\right)^{1/2}
\end{eqnarray}
}
\newline 
$\underline{\phi \phi^\ast \rightarrow \overline{f}f}:$
{\small 
\begin{eqnarray}
  \frac{c_f}{4 \pi  v^2 \Lambda^2} 
\frac{\lambda_f^2 d_{12}^2 m_\phi^4}{\left[\left(4m_\phi^2 
- m_h^2\right)^2 + m_h^2 \Gamma_h^2\right]}
  \left(1 - \frac{m_f^2}{m_\phi^2}\right)^{3/2}
\end{eqnarray}
}
Here the fermion Yukawa coupling is $\lambda_{f} = m_f/v$ where $v
\simeq 246$ GeV and $m_f$ is the fermion mass, while $c_f=1,3$ for
leptons and quarks respectively.  The contributions from the photon
mediated annihilations from the charge-radius operator are negligible
and we do not include these.

We implement the Lagrangian in Eq.~(\ref{Basic Lagrangian}) in {\tt
  CalcHEP} \cite{Pukhov:2004ca} and in {\tt MicrOMEGAs}
\cite{Belanger:2008sj} (using the {\tt LanHEP} module
\cite{Semenov:2008jy} to check the above implementation), in order to
compute the full $2 \to 2$ annihilation cross-section including finite
widths and to study the collider phenomenology.

\end{document}